\documentclass[aps]{revtex4}
\usepackage{amsbsy}
\usepackage{graphicx,color}
\usepackage{amsfonts}
\usepackage{amsmath}
\usepackage{amssymb}

\newcommand{\ds}{ _{\downarrow}}
\newcommand{\us}{ _{\uparrow}}
\newcommand{\up}{\uparrow}
\newcommand{\down}{\downarrow}

\begin{document}
\title{Dimer-dimer scattering length for fermions with different masses: analytical study for large mass ratio}
\author{F. Alzetto$^{(a)}$, R. Combescot$^{(a),(b)}$ and  X. Leyronas$^{(a)}$}
\address{(a) Laboratoire de Physique Statistique, Ecole Normale Sup\'erieure, UPMC  
Paris 06, Universit\'e Paris Diderot, CNRS, 24 rue Lhomond, 75005 Paris,  
France.}
\address{(b) Institut Universitaire de France, Paris, France}
\date{Received \today}

\begin{abstract}
We study the dimer-dimer scattering length $a_4$ for a two-component Fermi mixture in which the different fermions
have different masses $m\us$ and $m\ds$. This is made in the framework of the exact field theoretical method. 
In the large mass ratio domain the equations are simplified enough to lead to an analytical solution. In particular
we link $a_4$ to the fermion-dimer scattering length $a_3$ for the same fermions, and obtain 
the very simple relation $a_4=a_3/2$. The result $a_4 \simeq a_3/2$ is actually valid whatever the mass ratio
with quite good precision. As a result we find an analytical expression providing $a_4$ with a fairly good precision 
for any masses. To dominant orders for
large mass ratio it agrees with the literature. We show that, in this large mass ratio domain, the dominant
processes are the repeated dimer-dimer Born scatterings, considered earlier by Pieri and Strinati. We conclude
that their approximation, of retaining only these processes, is a fairly good one whatever the mass ratio.
\end{abstract}
\maketitle
\section{INTRODUCTION}

Ultracold atoms are a remarkable playground for a number of other fields of physics, such as condensed matter physics,
nuclear physics and astrophysics. This is due to the simplicity at very low temperature of the effective interaction between atoms
and moreover to the experimental ability to choose almost at will the corresponding parameters \cite{gps}. The resulting
physical systems correspond often to simple limiting situations of high interest in these other fields. The experimental realization
of the BEC-BCS crossover in Fermi gases is a wonderful example of the flexibility available in cold gases. As a result of the ability to vary
the effective interaction in a very wide range, it has been possible to go from the weakly attractive regime between fermions,
where at low temperature a BCS-like condensate arises, to the strongly attractive regime where molecules, or dimers, form
and these composite bosons undergo the standard Bose-Einstein condensation at low temperature. This crossover has provided
an explicit demonstration of the deep-seated connection between these two kinds of superfluids, which had been realized long ago
\cite{popov,kk,eagles,leg,nsr}. At the same time it displays, at unitarity where dimers are just appearing, a somewhat new example
of superfluidity.

When the interaction strength is just beyond the one corresponding to unitarity so that dimers are just forming, 
the physical situation is quite complex since the dimer size is very large and they overlap very strongly in the gas
leading to a complicated many-body problem, at higher temperature in the normal state as well as in the superfluid
state found at lower temperature. However when the interaction strength is increased toward the strong coupling
regime, the dimer size becomes very small and their overlap becomes negligible. In this case the composite nature
of the dimers becomes irrelevant and physically one deals with a simple boson gas. Just as for ultracold Bose gas,
the only relevant parameter to describe the low temperature physics is the boson-boson scattering length, in the
present case the dimer-dimer scattering length $a_4$. Hence knowing the precise value of this parameter, in terms of
the scattering length $a$ for fermions making up the dimer, is of utmost importance for the description of this strong coupling
limit.

This problem was first addressed by Haussmann \cite{haus} and by S\'a de Melo, Randeria and Engelbrecht \cite{sdm}
by different methods, which turn out to be equivalent to the Born approximation for this scattering problem. The corresponding
result is $a_4=2\,a$. This result was much improved by Pieri and Strinati \cite{ps} who considered repeated dimer-dimer 
scattering, instead of a single scattering corresponding to the Born approximation. They proceeded to sum up 
the corresponding series and obtained numerically $a_4 \simeq 0.75\,a$. Finally an exact treatment was given
by Petrov, Salomon and Shlyapnikov \cite{pss1,pss} who provided the numerical solution of the corresponding four-body
Schr\"odinger equation. This led them $a_4 \simeq 0.60\,a$. This problem was then taken up by
Brodsky, Klaptsov, Kagan, Combescot and Leyronas \cite{bkkcl} who gave an exact treatment of the same problem,
by making use of field theoretical methods. The numerical solution of their equations gave naturally the same result
$a_4 \simeq 0.60\,a$ as the one obtained by Petrov, Salomon and Shlyapnikov.

In view of the interest in fermionic mixtures made of different elements, such as $^6$Li -$^{40}$K mixtures,
Petrov, Salomon and Shlyapnikov extended their treatment \cite{pss2} to the case where the fermions making up
the dimers have different masses. The corresponding extension of the exact field theoretical treatment
was provided by Iskin and S\'a de Melo \cite{ism}, who provided numerical results for several mixtures of
specific interest. This method has been extended recently by Levinsen and Petrov \cite{lp} to the case of
narrow Feshbach resonances, aiming specifically at $^6$Li -$^{40}$K mixtures.
Here, in the case of a wide Feshbach resonance, we take over this technique to investigate the large mass ratio analytically.
Our aim is the same as in our recent work on the fermion-dimer scattering length \cite{acl}, namely to gain some insight
in the dimer-dimer scattering problem which might be used in dealing with more complicated problems
where this scattering is a building block.

In the present case we succeed in obtaining such an insight. We indeed find that the Pieri and Strinati
\cite{ps} approximation is asymptotically correct for large mass ratio. Taken with the fact that their result for
$a_4$ is also a quite good approximation when the two different fermions have equal mass, we come to the
conclusion that their approximation is quite good for any mass ratio. This is an important simplification
in the dimer-dimer scattering problem since this means that we do not have to take into account intermediate
states where one of the dimer is broken. Except for the irreducible process corresponding to the Born contribution,
this means that in all the intermediate states in this scattering process we have to deal with unbroken dimers.

\section{Basic equations}

As it is quite often done, we will call spin-up and spin-down the two different fermions with respective mass $m\us$ and $m\ds$.
This is a convenient convention frequently used in cold gases, although there is in principle no link with the physical spin of the particles.
The dimer is a bound state of one spin-up and one spin-down fermion. Clearly the scattering length $a_4$
we are looking for is proportional to the spin-up spin-down scattering length $a$. For dimensional reasons
$a_4$ depends only on the mass ratio $r=m\ds/m\us$. Since exchanging the two fermions does not change the dimer,
the result is unchanged when $r$ is changed into $1/r$. Hence the large mass limit we are mostly interested in
can be seen equally as the $r \rightarrow \infty$ limit or the $r\rightarrow 0$ limit.

\begin{figure}
\centering
{\includegraphics[width=\linewidth]{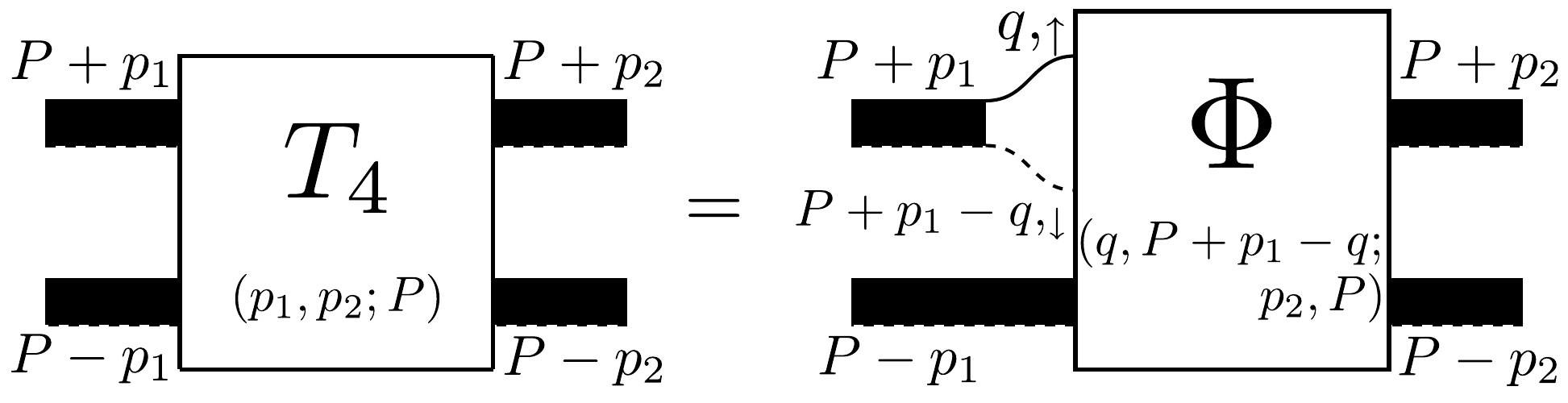}}
\caption{Diagrammatic representation of the relation Eq.(\ref{PhitoT4}) between the full dimer-dimer scattering vertex $T_4$ 
and the vertex $\Phi$. The black strips are for ladder diagrams corresponding to the dimer propagator $T_2$
given by Eq.(\ref{T_2}). The full line is for the $\up$-spin propagator. The dashed line is for the $\down$-spin propagator.}
\label{fig1}
\end{figure}

Let us first obtain the  generalization for different masses of the equations used by BKKCL \cite{bkkcl}.
The two basic vertices are again $T_4(p_1,p_2;P)$ and $\Phi(q_1,q_2;p_2, P)$. They are shown
in Fig.~\ref{fig1}. The first one describes the scattering of two dimers entering with respective four-momenta $P+p_1$ and $P-p_1$,
and outgoing with four-momenta $P+p_2$ and $P-p_2$.  In the second one the entering particles are two different
fermions with respective  four-momenta $q_1$ and $q_2$ and a dimer with four-momentum $P-q_1-q_2$, while
the outgoing particles are again two dimers with four-momenta $P+p_2$ and $P-p_2$. We will take the convention
for $\Phi$ that the first variable $q_1$ corresponds to the spin-up particle while the second one $q_2$ corresponds
to the spin-down one, explicitly $\Phi(q_1 \up,q_2\down;p_2, P)$. In contrast with the equal mass case
$\Phi(q_1,q_2;p_2, P)$ is no longer symmetric in the exchange of $q_1$ and $q_2$.

\begin{figure}
\centering
{\includegraphics[width=\linewidth]{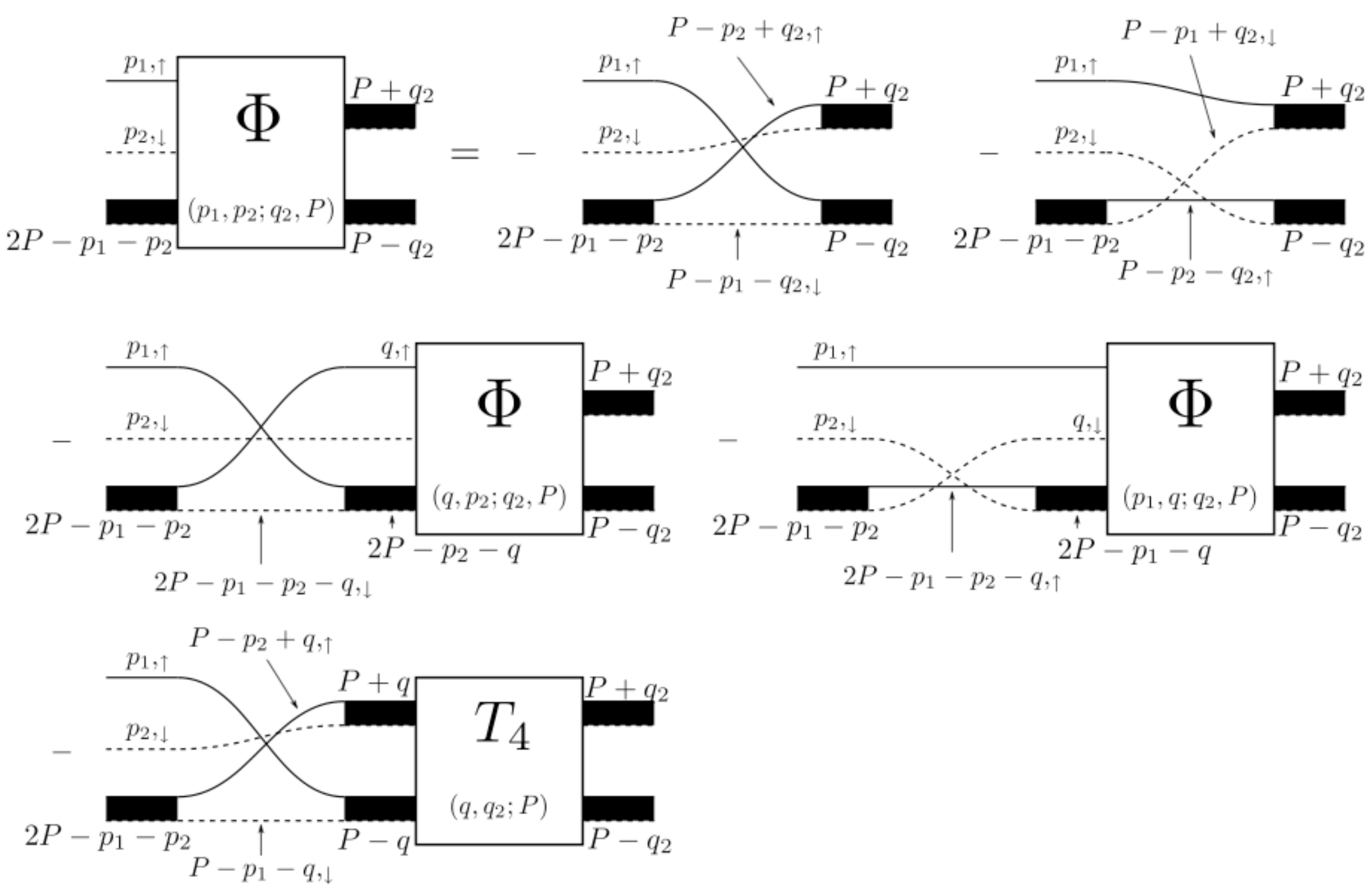}}
\caption{Diagrammatic representation of the equation Eq.(\ref{Phi})
    for the vertex $\Phi$. The black strips are for ladder diagrams corresponding to the dimer propagator $T_2$
given by Eq.(\ref{T_2}). Full lines are for $\up$-spin propagators. Dashed lines are for $\down$-spin propagators.}
\label{fig2}
\end{figure}

It is clear that the first process in $T_4$ involves the interaction of a fermion of one dimer with a fermion of the other
dimer. Hence one has first to split open a dimer line into its fermionic components, all the subsequent processes
being described by $\Phi$. This leads to the equality shown in Fig.~1. Algebraically it reads:
\begin{equation}\label{PhitoT4}
T_4(p_1,p_2;P)=\sum_{k}G\us(k)G\ds(P+p_1-k)\Phi(k,P+p_1-k;p_2, P)
\end{equation}
where $\sum\limits_k\equiv i \int d{\bf k}d\omega/(2\pi)^4$ is for the summation over momentum and energy
and the propagators are given by $G_{\uparrow, \downarrow }(k)=[\omega -k^2/2m_{\uparrow, \downarrow } +i\epsilon ]^{-1}$,
with $\epsilon \to 0^+$.
In turn we can write an equation for $\Phi$, in the same spirit as a Bethe-Salpeter equation. The first process is
a break up of the entering dimer, in order to allow the interaction of one of the free fermions with a fermion from the dimer.
Actually other interactions between these same fermions may also occur after the first one and the resummation
of all these processes leads merely to a dimer propagator $T_2$. Afterwards any process may follow, which is again
depicted by $\Phi$. However one must have in mind that by definition, in $\Phi$, the first process can not be an interaction
between the two free fermions (otherwise one would merely have an entering dimer, instead of two free fermions, which is
already accounted for by the entering dimer propagator in $T_4$). But such a process is quite allowed after the first
interaction, and since they are not accounted for by $\Phi$ we have to add terms to describe them. The repeated
interaction of the free fermions leads to another dimer propagator, so at this stage we have two dimers. This can
be taken as the outgoing state, and the corresponding diagrams correspond merely to the Born term for the
diagrammatic expansion for $\Phi$. However it is also possible to have any other process after these two dimers
have been formed, which is fully described by $T_4$. For each of the process we have described, we have
actually two possibilities depending on the spin of the involved particles. However this is not true for the last
$T_4$ term we have just described, since one sees easily that exchanging the spins is equivalent to a change
in dummy variables. This leads finally to the diagrammatic equation depicted in Fig.~\ref{fig2}, which reads algebraically:
\begin{multline}\label{Phi}
    \Phi(q_1,q_2;p_2, P)=-G\us(P-q_2-p_2)G\ds(P-q_1+p_2)-G\ds(P-q_1-p_2)G\us(P-q_2+p_2) \\
   - \sum_{k}G\us(k)G\ds(2P-q_1-q_2-k)T_2(2P-q_2-k)\Phi(k,q_2;p_2, P) -  \sum_{k}G\ds(k)G\us(2P-q_1-q_2-k)T_2(2P-q_1-k)\Phi(q_1,k;p_2, P) \\
    -\sum_{Q}G\us(Q-q_2)G\ds(2P-Q-q_1) T_2(2P-Q)T_2(Q)T_4(P-Q,p_2;P)
\end{multline}
The dimer propagator is obtained as usual by summing up the ladder diagrams and is given by:
\begin{equation}\label{T_2}
T_2(P) = \frac{2\pi}{\mu}\,
    \frac{1}{a^{-1}-\sqrt{2\mu(\mathbf{P}^2/2M-E)}} \equiv T_2(\mathbf{P},E)
\end{equation}
for the four-momentum $P = \{\textbf{P},E\}$. Here $\mu $ is the reduced mass $\mu =m\us m\ds/M=m\us r/(1+r)$, while
$M$ is the total mass $M=m\us+m\ds=m\us (1+r)$.

Just as in \cite{bkkcl} the scattering length is obtained directly from $T_4$, evaluated at the dimer binding energy
$E_b=1/(2\mu a^2)$, the only difference being that we  have to use the dimer reduced mass $\mu $ and the reduced 
mass $M/2$ of two dimers:
\begin{equation}\label{a4}
\left(\frac{2\pi}{\mu ^2a}\right)^2T_4(0,0;\{{\bf 0},-E_b\})=\frac{2\pi(2a_4)}{M/2}.
\end{equation}
Since for the scattering length problem we have to take $p_2=0$ and $P=\{\textbf{0},-E_b\}$ we do not write anymore
explicitly these variables. Hence Eq.(\ref{a4}) becomes:
\begin{equation}\label{a4bar}
{\bar a}_4 \equiv \frac{a_4}{a}=\frac{\pi}{2}\,\frac{(1+r)^5}{r^4}\frac{T_4(0)}{(a m\us)^3}
\end{equation}
In the following we take for convenience $a$ and $m\us$ as unit for length and mass, i.e. we set $a=1$ and $m\us=1$.
This last step breaks apparently the invariance of ${\bar a}_4$ under $r \rightarrow 1/r$, but this symmetry is
naturally still satisfied in the final results.

A very important simplification is that, just as in \cite{bkkcl}, the calculation of the two $\Phi$ terms in the right-hand side of
Eq.(\ref{Phi}) requires only the knowledge of the "on the shell" value for $\Phi$ with respect to variable $k$. This allows
to consider Eq.(\ref{Phi}) only for "on the shell" values for variables $q_1$ and $q_2$. We denote $\Phi({\bf q}_1,{\bf q}_2)$
the corresponding "on the shell" value of $\Phi$ (it is not symmetric under the exchange of ${\bf q}_1$ and ${\bf q}_2$). 
Similarly when Eq.(\ref{Phi}) is inserted into Eq.(\ref{PhitoT4}), the resulting 
equation requires only the knowledge of $\Phi({\bf q}_1,{\bf q}_2)$. Finally it is again convenient to continue the equations
to purely imaginary values for the frequencies. The corresponding value for $T_4(Q)=T_4(\{\textbf{Q},i \Omega \}) 
\equiv t_4(\textbf{Q},\Omega)$ is real and the equations contain only real quantities.

Let us now give the various terms entering the equations for $\Phi$ and $t_4$. The first two terms in Eq.(\ref{Phi})
correspond to the Born approximation and give to $\Phi({\bf q}_1,{\bf q}_2)$ a contribution:
\begin{eqnarray}\label{phiborn}
\Phi_B({\bf q}_1,{\bf q}_2)=-8 \mu ^2 \,\frac{1}{1+q_1^2}\, \frac{1}{1+q_2^2}
\end{eqnarray}
where now $q_1=|{\bf q}_1|$ and $\mu =r/(1+r)$. The two next terms in Eq.(\ref{Phi}) give two contributions:
\begin{eqnarray}\label{phiphi1}
\Phi_{\Phi 1}({\bf q}_1,{\bf q}_2)=\frac{2}{(2 \pi )^3}  \int d{\bf k}\; \frac{
T_2(-1/\mu-q_1^2/2-k^2/2r ,{\bf q}_1+{\bf k})\,\Phi({\bf q}_1,{\bf k})}
{2/\mu +q_1^2+q_2^2/r+k^2/r+({\bf q}_1+{\bf q}_2+{\bf k})^2}
\end{eqnarray}
where the integration over the azimuthal angle of ${\bf k}$ with respect to ${\bf q}_1$
can be explicitly performed, since $\Phi({\bf q}_1,{\bf k})$ as well as $T_2$ depend only
on the polar angle between ${\bf k}$ and ${\bf q}_1$. This leaves a double integration to be performed.
The other contribution is
\begin{eqnarray}\label{phiphi2}
\Phi_{\Phi 2}({\bf q}_1,{\bf q}_2)=\frac{2}{(2 \pi )^3}  \int d{\bf k}\; \frac{
T_2(-1/\mu-q_2^2/2r-k^2/2 ,{\bf q}_2+{\bf k})\,\Phi({\bf k},{\bf q}_2)}
{2/\mu +q_1^2+q_2^2/r+k^2+({\bf q}_1+{\bf q}_2+{\bf k})^2/r}
\end{eqnarray}
where the integration over the azimuthal angle of ${\bf k}$ with respect to ${\bf q}_2$
can again be explicitly performed leaving again a double integral.
Finally the last term gives, after the change of variable $Q \rightarrow P-Q$:
\begin{eqnarray}\label{phit4}
\Phi_T({\bf q}_1,{\bf q}_2)=\frac{4}{(2 \pi )^4}  \int d{\bf Q} \int_{-\infty}^{\infty}\!\!d\Omega \; \frac{
|T_2\left(-1/2\mu-i\Omega ,{\bf Q}\right)|^2\,t_4\left(\textbf{Q},\Omega \right)}
{\left[1/\mu -2i\Omega +q_1^2+({\bf Q}-{\bf q}_1)^2/r\right]
\left[1/\mu +2i\Omega +q_2^2/r+({\bf Q}+{\bf q}_2)^2\right]}
\end{eqnarray}
Since $t_4\left(\textbf{Q},\Omega \right)$ as well as $T_2$ depend only on the modulus of ${\bf Q}$,
the angular average over ${\bf Q}$ can be explicitly performed by making use of:
\begin{eqnarray}\label{formula}
 \int\frac{d\Omega_k }{4\pi }\frac{1}{a+{\bf k}.{\bf u}}\,\frac{1}{b+{\bf k}.{\bf v}}=\frac{1}{2A}
 \ln\frac{ab-k^2 {\bf u}.{\bf v}+A}{ab-k^2 {\bf u}.{\bf v}-A}
\end{eqnarray}
where
\begin{eqnarray}\label{gdA}
A=\sqrt{(akv)^2+(bku)^2-2abk^2{\bf u}.{\bf v}-k^4\left[(uv)^2-({\bf u}.{\bf v})^2\right]}
\end{eqnarray}
This leaves again a double integral to be performed.
To summarize we have explicitly:
\begin{eqnarray}\label{phitot}
\Phi({\bf q}_1,{\bf q}_2)=\Phi_B({\bf q}_1,{\bf q}_2)+\Phi_{\Phi 1}({\bf q}_1,{\bf q}_2)+\Phi_{\Phi 2}({\bf q}_1,{\bf q}_2)
+\Phi_T({\bf q}_1,{\bf q}_2)
\end{eqnarray}
where the four terms in the right-hand side are defined respectively by Eq.(\ref{phiborn}),Eq.(\ref{phiphi1}),Eq.(\ref{phiphi2}) and Eq.(\ref{phit4}).

Similarly, substituting Eq.(\ref{Phi}) in Eq.(\ref{PhitoT4}) we obtain for $t_4$:
\begin{eqnarray}\label{t4tot}
t_4\left(\textbf{q},\nu \right)=T_B(\textbf{q},\nu)+T_{\Phi}(\textbf{q},\nu)+T_T(\textbf{q},\nu)
\end{eqnarray}
After the same change of variable $Q \rightarrow P-Q$ the last term reads:
\begin{eqnarray}\label{t4t4}
T_T(\textbf{q},\nu)=\frac{1}{(2 \pi )^4}  \int d{\bf Q} \int_{-\infty}^{\infty}\!\!d\Omega  \; 
K(\textbf{q},\nu;\textbf{Q},\Omega)\,
|T_2\left(-1/2\mu-i\Omega ,{\bf Q}\right)|^2\,t_4\left(\textbf{Q},\Omega \right)
\end{eqnarray}
where the symmetric kernel $K(\textbf{q},\nu;\textbf{Q},\Omega)=K(\textbf{Q},\Omega;\textbf{q},\nu)$ is given by:
\begin{eqnarray}\label{K}
K(\textbf{q},\nu;\textbf{Q},\Omega)=\sum_{k}G\us(k)G\ds(P+q-k)G\us(k-q-Q)G\ds(P+Q-k)
\end{eqnarray}
with $q = \{\textbf{q},i\nu\}$. This can be written in a more symmetric way by the change of variable $k \rightarrow k+(P+q+Q)/2$:
\begin{eqnarray}\label{Kbis}
K(\textbf{q},\nu;\textbf{Q},\Omega)=\sum_{k}G\us(k+\frac{P+q+Q}{2})G\ds(\frac{P+q-Q}{2}-k)
G\us(k+\frac{P-q-Q}{2})G\ds(\frac{(P-q+Q}{2}-k)
\end{eqnarray}
After performing in Eq.(\ref{K}) the frequency integration over $\omega $ 
and making the change ${\bf k} \rightarrow {\bf k}+{\bf q}$, one is left with expressions which can be written 
in terms of products of two rational functions, each one being
of the form $1/(a+{\bf k}.{\bf u})$. The angular average can be performed by making use of Eq.(\ref{formula}).
One is left with a simple integration over the modulus of ${\bf k}$. This means that evaluation of Eq.(\ref{t4t4})
requires four integrations. However we will not write the lengthy
resulting expressions, which are only necessary to perform the full numerical calculation for the general
case of two different masses $m\us$ and $m\ds$.

Then the Born contribution $T_B(\textbf{q},\nu)$ is merely given by:
\begin{eqnarray}\label{t4B}
T_B(\textbf{q},\nu)=-2K(\textbf{q},\nu;\textbf{0},0)
\end{eqnarray}

Finally the last term $T_{\Phi}(\textbf{q},\nu)$ is given by:
\begin{eqnarray}\label{t4Phi}
T_{\Phi}(\textbf{q},\nu)= -\frac{4}{(2\pi )^6} \int d{\bf k}\,d{\bf k}'
\frac{T_2\left(-1/\mu-k^2/2-k'^2/2r ,{\bf k}+{\bf k}'\right)\,\Phi({\bf k},{\bf k}')}
{\left[1/\mu -2i\nu +k^2+({\bf k}-{\bf q})^2/r\right]
\left[1/\mu +2i\nu +k'^2/r+({\bf k}'+{\bf q})^2\right]}\;+ (\;q\;\rightarrow -q\;)
\end{eqnarray}
Since the result depends clearly only on the modulus of ${\bf q}$ and the dependence on ${\bf q}$ is explicit, 
the integrand can be angularly averaged over the direction of ${\bf q}$ by making use of Eq.(\ref{formula}).
This leaves a triple integral over the moduli of ${\bf k}$ and ${\bf k}'$, and the angle between them.

In order to solve numerically Eq.(\ref{phitot}) and Eq.(\ref{t4tot}) for $\Phi({\bf q}_1,{\bf q}_2)$ and 
$t_4\left(\textbf{q},\nu \right)$ and then obtain $a_4$ from Eq.(\ref{a4}), we have discretized these equations.
This leads to a set of linear equations, the solution corresponding to a matrix inversion.
This has been performed making use of the standard LAPACK routines. This works quite well for not too high mass
ratios, and we estimate the precision of our results to be typically 2\%. However for mass ratio typically above 500,
the numerics becomes unreliable. This is easily understood from the somewhat singular features which emerge
from our anaytical solution, presented below, for very large mass ratio. We note finally that we never find any zero
eigenvalue for the matrix to be inverted. This means that the bound states discussed in \cite{pss2} do not play any
role in the calculation of $a_4$.

\section{Sum rule}

In a way completely analogous to what we have found in our study of the atom-dimer scattering length \cite{acl}, a quite useful sum rule
can be obtained by analyzing the convergence of the various integrals when the variables go to infinity. It is natural to assume
that the solutions $\Phi({\bf q}_1,{\bf q}_2)$ and $t_4\left(\textbf{q},\nu \right)$ we are looking for have physical ranges
corresponding to their variables, and that beyond these ranges these functions go rapidly enough to zero for the various integrals
to be convergent. This assumption is confirmed by
our numerical calculations. This property allows to study the behaviour of $\Phi({\bf q}_1,{\bf q}_2)$ and $t_4\left(\textbf{q},\nu \right)$
for large values of the variables.

Let us first consider the $t_4$ equation Eq.(\ref{t4tot}) and begin by the Born contribution $T_B(\textbf{q},\nu)$ given by Eq.(\ref{t4B}) and Eq.(\ref{K}). 
Since we are interested in the case where $\textbf{q}$ and $\nu$ are large, $P$ is negligible in this regime and we are left with:
\begin{eqnarray}\label{t4Blarge}
T_B(\textbf{q},\nu) \approx -2\sum_{k}G\us(k)G\ds(q-k)G\us(k-q)G\ds(-k)
\end{eqnarray}
It is then convenient to consider more appropriate energy variables. Since energy is homogeneous to momentum squared, we set
$\nu=q_{\nu}^2$ and similarly for the integration variable $\omega =k_{\omega }^2$. For large values of $\textbf{q}$ and $\nu$,
the natural variable to consider is $\rho=\sqrt{{\bf q}^2+q_{\nu}^2}$, corresponding to introduce a radial coordinate for $|{\bf q}|$ and $q_{\nu}$.
Similarly we can introduce $r=\sqrt{{\bf k}^2+k_{\omega }^2}$ (not to be confused with the mass ratio). Now for example $G\us(k)$ is homogeneous to $r^{-2}$, and similarly
for the other Green's functions entering Eq.(\ref{t4Blarge}). Hence for homogeneity reasons we have $T_B(\textbf{q},\nu) \sim \rho^{-3}$
since the summation $\sum_{k}$ introduces a factor homogeneous to $r^5$. More precisely, making the change of variable $r={\bar r}\rho$
allows to obtain in Eq.(\ref{t4Blarge}) a prefactor $\rho^{-3}$, the remaining factor being independent of $\rho$. One can naturally write
more explicit expressions for $T_B(\textbf{q},\nu)$, although it is complicated to go to a full analytical result. However we will not
need these expressions and they merely confirm our result that $T_B(\textbf{q},\nu) \sim ({\bf q}^2+q_{\nu}^2)^{-3/2}$ which comes simply
from our homogeneity analysis.

We can now make a similar analysis for the term $T_T(\textbf{q},\nu)$ in the $t_4$ equation. Indeed since in Eq.(\ref{t4t4})
we are interested in large values of $\textbf{q}$ and $\nu$, while the integration variables ${\bf Q}$ and $\Omega $ are
effectively bounded by the factor $t_4(\textbf{Q},\Omega)$, the kernel reduces in this limit to:
\begin{eqnarray}\label{}
K(\textbf{q},\nu;\textbf{Q},\Omega) \approx K(\textbf{q},\nu;\textbf{0},0)
\end{eqnarray}
as it is quite clear from the symmetric form Eq.(\ref{Kbis}) for $K$. Hence this kernel factorizes out and we are left with:
\begin{eqnarray}\label{t4t4large}
T_T(\textbf{q},\nu)=\frac{1}{(2 \pi )^4} K(\textbf{q},\nu;\textbf{0},0) \int d{\bf Q} \int_{-\infty}^{\infty}\!\!d\Omega \;
|T_2\left(-1/2\mu-i\Omega ,{\bf Q}\right)|^2\,t_4\left(\textbf{Q},\Omega \right)
\end{eqnarray}
which has naturally the same behaviour as the Born term, namely $T_T(\textbf{q},\nu) \sim ({\bf q}^2+q_{\nu}^2)^{-3/2}=\rho^{-3}$.

Finally we analyze in the same way the last term $T_{\Phi}(\textbf{q},\nu)$ in this $t_4$ equation from its explicit expression Eq.(\ref{t4Phi}). 
With the ${\bf k}$ and ${\bf k}'$ variables bounded by the $\Phi({\bf k},{\bf k}')$ factor, the $\textbf{q}$ and $\nu$ dependence comes explicitly 
from the denominators and is given by $T_{\Phi}(\textbf{q},\nu) \sim ({\bf q}^2+q_{\nu}^2)^{-2}=\rho^{-4}$. Hence it converges toward zero faster than
the two other terms and the overall behaviour of $t_4(\textbf{q},\nu)$ is apparently $t_4(\textbf{q},\nu) \sim \rho^{-3}$.

However when we insert this behaviour in the integral factor found in Eq.(\ref{t4t4large}) for the large $({\bf q}^2+q_{\nu}^2)$ behaviour we find
a discrepancy. Indeed from Eq.(\ref{T_2}), and for large ${\bf Q}$ and $\Omega$, we have  $|T_2\left(-1/2\mu-i\Omega ,{\bf Q}\right)|^2 \sim R^{-2}$,
where we have again made the change $\Omega =Q^2_{\Omega }$ for the energy variable and introduced the radial variable 
$R=\sqrt{{\bf Q}^2+Q_{\Omega }^2}$. This leads to $|T_2\left(-1/2\mu-i\Omega ,{\bf Q}\right)|^2\,t_4\left(\textbf{Q},\Omega \right) \sim R^{-5}$.
However we have $d{\bf Q}\,d\Omega \sim R^4\,dR$ so that the integral in Eq.(\ref{t4t4large}) diverges as $dR/R$. This is in
contradiction with our initial assumption that $t_4\left(\textbf{Q},\Omega \right)$ insures the convergence of the integral.

The only escape is that the contributions from the Born term and from the $T_T$, which have exactly the same power law dependence,
cancel out so that the actual decrease of $t_4\left(\textbf{Q},\Omega \right)$ is faster than $R^{-3}$. This occurs if the corresponding
coefficients cancel exactly, which leads to the sum rule:
\begin{eqnarray}\label{sumrule}
\frac{1}{(2 \pi )^4}\int d{\bf Q} \int_{-\infty}^{\infty}\!\!d\Omega \; |T_2\left(-1/2\mu-i\Omega ,{\bf Q}\right)|^2\,t_4\left(\textbf{Q},\Omega \right)=2
\end{eqnarray}
Upon checking this result numerically, we have found that it is in very good agreement with our calculations.

Let us now analyze in the same way the $\Phi$ equation Eq.(\ref{phitot}). First the Born term is explicit and, for large ${\bf q}_1$ and ${\bf q}_2$,
it behaves as:
\begin{eqnarray}\label{phibornlarge}
\Phi_B({\bf q}_1,{\bf q}_2) \approx -8 \mu ^2 \,\frac{1}{q_1^2q_2^2}
\end{eqnarray}
Then we consider the $\Phi_T({\bf q}_1,{\bf q}_2)$ term given by Eq.(\ref{phit4}).
Again the factor $t_4\left(\textbf{Q},\Omega \right)$, assumed to go rapidly enough to zero for large $\textbf{Q}$ and $\Omega$,
makes these variables $\textbf{Q}$ and $\Omega$ effectively bounded. Hence when we consider very large values of ${\bf q}_1$ and 
${\bf q}_2$, we can forget about $\textbf{Q}$ and $\Omega$ in the explicit denominators, and the product of these denominators is
given in this case by $q_1^2 (1+1/r)q_2^2(1/r+1)=(q_1 q_2/\mu )^2$. This leads for large $q_1$ and $q_2$ to:
\begin{eqnarray}\label{phit4large}
\Phi_T({\bf q}_1,{\bf q}_2) \approx \frac{4}{(2 \pi )^4} \frac{\mu ^2}{q_1^2 q_2^2} \int d{\bf Q} \int_{-\infty}^{\infty}\!\!d\Omega \; 
|T_2\left(-1/2\mu-i\Omega ,{\bf Q}\right)|^2\,t_4\left(\textbf{Q},\Omega \right)
\end{eqnarray}
We consider finally the $\Phi_{\Phi 1}$ and $\Phi_{\Phi 2}$ terms. Our assumption is that $\Phi({\bf q}_1,{\bf q}_2)$
goes rapidly enough to zero at large ${\bf q}_1$ and ${\bf q}_2$ for having the integrals in $\Phi_{\Phi 1}$ and $\Phi_{\Phi 2}$
converge for large ${\bf k}$ due to the $\Phi$ factor. This means explicitly that $ \int d{\bf k}\,\Phi({\bf q}_1,{\bf k})$ and
$ \int d{\bf k}\,\Phi({\bf k},{\bf q}_2)$ are convergent. However this is not true if we take the behaviour $\sim 1/(q_1^2 q_2^2)$
produced by the Born term and the $\Phi_T$ term, since this gives an integral $ \int d{\bf k}/k^2 \sim  \int d{k}$, which is
divergent. If the dominant behaviour was coming from the $\Phi_{\Phi}$ terms themselves, then $\Phi({\bf q}_1,{\bf q}_2)$
would go even more slowly to zero at infinity and the situation would be even worst (one can actually check that this case
does not arise). Hence we have again a contradiction with our starting hypothesis, but there is also the same way out,
namely that the contribution of the Born term and the $\Phi_T$ cancel precisely, which leads to a decrease of 
$\Phi({\bf q}_1,{\bf q}_2)$ faster than $\sim 1/(q_1^2 q_2^2)$ for large ${\bf q}_1$ and ${\bf q}_2$. We see from
Eq.(\ref{phibornlarge}) and Eq.(\ref{phit4large}) that the condition for this to happen is exactly the sum rule already 
found Eq.(\ref{sumrule}). This result is not so surprising when we remember that the $t_4$ equation is obtained by carrying
the $\Phi$ equation into Eq.(\ref{PhitoT4}).

We note finally that, in our work on the atom-dimer scattering length \cite{acl}, we have shown that the sum rule we have found
in this case is merely a direct consequence of the Pauli principle, namely the fact that two identical fermions can not be found at the same place.
In the present case such a physical interpretation is less obvious since the frequency dependence of
$t_4\left(\textbf{Q},\Omega \right)$ enters. Anyway we have not tried to find a specific physical interpretation.

\section{Very heavy mass equations}\label{vhma}

Let us now consider how the preceding equations simplify when we consider the limiting case where the two masses
$m\us$ and $m\ds$ are very different. We can equivalently consider that the mass ratio $r=m\ds/m\us$ goes to zero
or to infinity. We will take this last option in the following since it is somewhat more convenient with the asymmetric
notations we have chosen. This implies $\mu \rightarrow 1$ in this limit and the Born term Eq.(\ref{phiborn}) in the 
$\Phi$ equation becomes:
\begin{eqnarray}\label{phiborninf}
\Phi_B^{\infty}({\bf q}_1,{\bf q}_2)=-8  \,\frac{1}{1+q_1^2}\, \frac{1}{1+q_2^2}
\end{eqnarray}
In the $\Phi_{\Phi 1}$ and $\Phi_{\Phi 2}$ terms we may replace in Eq.(\ref{T_2}) $T_2(E,\mathbf{P})$ by $2\pi /(1-\sqrt{-2E})$
since $M \rightarrow \infty$ in this limit. This leads to:
\begin{eqnarray}\label{phiphi1inf}
\Phi_{\Phi 1}^{\infty}({\bf q}_1,{\bf q}_2)=-\frac{1}{2 \pi ^2} \frac{1}{\sqrt{2+q_1^2}-1} \int d{\bf k}\; \frac{
\,\Phi^{\infty}({\bf q}_1,{\bf k})}
{2 +q_1^2+({\bf q}_1+{\bf q}_2+{\bf k})^2}
\end{eqnarray}
and
\begin{eqnarray}\label{phiphi2inf}
\Phi_{\Phi 2}^{\infty}({\bf q}_1,{\bf q}_2)=-\frac{1}{2 \pi ^2}  \int d{\bf k}\; \frac{1}{\sqrt{2+k^2}-1}
\;\frac{\Phi^{\infty}({\bf k},{\bf q}_2)}{2 +q_1^2+k^2}
\end{eqnarray}
Let us finally consider the $\Phi_T$ term which gives rise to more problems. If we proceed in the same way as above, we find:
\begin{eqnarray}\label{phit4inf}
\Phi_T^{\infty}({\bf q}_1,{\bf q}_2)=\frac{4}{(2 \pi )^2}  \int d{\bf Q} \int_{-\infty}^{\infty}\!\!d\Omega \; \frac{1}
{\left[1 -2i\Omega +q_1^2\right]
\left[1 +2i\Omega +({\bf Q}+{\bf q}_2)^2\right]}\;\frac{t_4^{\infty}\left(\textbf{Q},\Omega \right)}{|\sqrt{1+2i\Omega }-1|^2}
\end{eqnarray}
However we see that the $\Omega $ integration diverges as $ \int\,d\Omega/\Omega^2$ for $\Omega \rightarrow 0$, since
$|\sqrt{1+2i\Omega }-1|^2 \simeq 4\Omega ^2$ in this limit. Indeed the explicit denominators go to finite values
and there is no reason to have $t_4(\textbf{Q},\Omega=0)=0$. This can anyway be checked numerically. In particular this
does not happen for ${\bf Q}={\bf 0}$ since, in the large $r$ limit we have from Eq.(\ref{a4bar}) (with our reduced variables):
\begin{equation}\label{a4barinf}
{\bar a}_4=\frac{\pi}{2}\,r\,t_4({\bf 0},0)
\end{equation}
The existence of this divergence shows that we have handled the $|T_2|^2$ term too rapidly, since it is at the origin of the
divergence which does not exist naturally in the general equation Eq.(\ref{phit4}). We must keep $M \approx r$ without
setting immediately $1/M=0$. In this case we have to write:
\begin{equation}\label{T_2inf}
T_2(E,\mathbf{P}) = \frac{2\pi}{1-\sqrt{\mathbf{P}^2/r-2E}}
\end{equation}
which leads, instead of Eq.(\ref{phit4inf}), to:
\begin{eqnarray}\label{phit4inf1}
\Phi_T^{\infty}({\bf q}_1,{\bf q}_2)=\frac{4}{(2 \pi )^2}  \int d{\bf Q} \int_{-\infty}^{\infty}\!\!d\Omega \; \frac{t_4^{\infty}\left(\textbf{Q},\Omega \right)}
{\left[1 -2i\Omega +q_1^2\right]
\left[1 +2i\Omega +({\bf Q}+{\bf q}_2)^2\right]}\frac{|\sqrt{1+2i\Omega +{\bf Q}^2/r}+1|^2}{4\Omega ^2+{\bf Q}^4/r^2}
\end{eqnarray}
We see that the divergence has disappeared. Instead we find a factor $1/(4\Omega ^2+{\bf Q}^4/r^2)$ which, in the limit
$r \rightarrow \infty$, is strongly peaked around $\Omega =0$ and acts in practice as a $\delta$ function:
\begin{eqnarray}\label{deltaf}
\frac{1}{4\Omega ^2+{\bf Q}^4/r^2} \approx \frac{\pi r}{2{\bf Q}^2}\,\delta(\Omega )
\end{eqnarray}
As a result only $t_4(\textbf{Q},\Omega=0)$ appears in the $\Phi$ equation so that we may write the equation for $t_4$
only for $\Omega =0$, which is a quite convenient simplification. Taking into account that $t_4(\textbf{Q},\Omega)$ depends
only on the modulus $Q=|{\bf Q}|$, we set:
\begin{eqnarray}\label{tbar}
{\bar t}_4(Q)=r\,t_4(\textbf{Q},\Omega=0)
\end{eqnarray}
which is just the quantity coming in the scattering length we are looking for:
\begin{equation}\label{a4barinf1}
{\bar a}_4 =\frac{\pi}{2}\,{\bar t}_4(0)
\end{equation}
Taking the limit $r \rightarrow \infty$ in the numerator of Eq.(\ref{phit4inf1}) and performing the angular average over the direction
of ${\bf Q}$, we finally end up with:
\begin{eqnarray}\label{phit4inf2}
\Phi_T^{\infty}({\bf q}_1,{\bf q}_2)=\frac{2}{(1+q_1^2)\,q_2}   \int_{0}^{\infty} dQ\, \frac{{\bar t}_4^{\infty}
(Q)}{Q}\,\ln\frac{1+(Q+q_2)^2}{1+(Q-q_2)^2}
\end{eqnarray}

We can now write the $t_4$ equation Eq.(\ref{t4tot}) with the variable $\nu$ set to zero, which is the only thing we need as we have
just seen. The Born contribution is given by:
\begin{eqnarray}\label{t4Binf}
T_B(\textbf{q},0)=-2K(\textbf{q},0;\textbf{0},0)
\end{eqnarray}
where $K(\textbf{q},0;\textbf{0},0)$ is obtained from Eq.(\ref{K}). Since in the limit $m\ds \rightarrow \infty$, $G\ds(k)$ reduces
to $[\omega +i\epsilon ]^{-1}$ the calculation is fairly simple and leads to:
\begin{eqnarray}\label{t4Binf1}
T_B^{\infty}(\textbf{q},0) \equiv t_B^{\infty}(q)=\frac{4}{\pi }\,\frac{1}{q^2+4}
\end{eqnarray}
The $T_{\Phi}$ term is also easily obtained in the limit $r \rightarrow \infty$ from Eq.(\ref{t4Phi}), making use of the limiting expression
for $T_2$ and performing the angular average over the direction of ${\bf q}$. This gives:
\begin{eqnarray}\label{t4Phiinf}
T_{\Phi}^{\infty}(\textbf{q},0)  \equiv t_{\Phi}^{\infty}(q) = \frac{1}{\pi ^3}\,\frac{1}{q} \int_{0}^{\infty} dk\,
\frac{k^2}{1+k^2}\,\frac{1}{\sqrt{2+k^2}-1} \int_{0}^{\infty} dk'\,k'\,\ln\frac{1+(k'+q)^2}{1+(k'-q)^2}
 \int \frac{d\Omega_{\bf k}' }{4\pi }\Phi^{\infty}({\bf k},{\bf k}')
 \end{eqnarray}
Finally in the last term of the $t_4$ equation, coming from Eq.(\ref{t4t4}), we meet the same troubles as in the last term
of the $\Phi$ equation if we use the simple limiting expression for $T_2$. In the same way as what we have
done to derive $\Phi_T^{\infty}({\bf q}_1,{\bf q}_2)$, we have to use Eq.(\ref{T_2inf}) for $T_2$. This leads in the same way to a
factor proportional to $\delta(\Omega )$ justifying the fact that we write the $t_4$ equation only for zero frequency.
Actually the last two factors in Eq.(\ref{t4t4}) are the same as those appearing in Eq.(\ref{phit4}). Hence the only
difference is that we have now to calculate the kernel $K(\textbf{q},0;\textbf{Q},0)$ for $r \rightarrow\infty$. This proceeds
just as for the Born contribution Eq.(\ref{t4Binf1}) and leads basically to the same result, provided ${\bf q}$ is replaced
by ${\bf q}+{\bf Q}$. This gives:
\begin{eqnarray}\label{}
K(\textbf{q},0;\textbf{Q},0)=-\frac{2}{\pi }\,\frac{1}{({\bf q}+{\bf Q})^2+4}
\end{eqnarray}
When the angular average over the direction of ${\bf q}$ is performed, just as in the preceding term, one finds:
\begin{eqnarray}\label{t4t4inf}
T_T^{\infty}(\textbf{q},0)  \equiv t_T^{\infty}(q)=-\frac{1}{\pi}\,\frac{1}{q}   \int_{0}^{\infty} dQ \,
\frac{{\bar t}_4^{\infty}(Q)}{Q}\,\ln\frac{4+(Q+q)^2}{4+(Q-q)^2}
\end{eqnarray}
To summarize the $\Phi$ equation becomes:
\begin{eqnarray}\label{phitotinf}
\Phi^{\infty}({\bf q}_1,{\bf q}_2)=\Phi_B^{\infty}({\bf q}_1,{\bf q}_2)+\Phi_{\Phi 1}^{\infty}({\bf q}_1,{\bf q}_2)+\Phi_{\Phi 2}^{\infty}({\bf q}_1,{\bf q}_2)
+\Phi_T^{\infty}({\bf q}_1,{\bf q}_2)
\end{eqnarray}
where the four terms in the right-hand side are defined respectively by Eq.(\ref{phiborninf}), Eq.(\ref{phiphi1inf}), Eq.(\ref{phiphi2inf}) 
and Eq.(\ref{phit4inf2}), while the $t_4$ equation reads in this limit, with the definition Eq.(\ref{tbar}):
\begin{eqnarray}\label{t4totinf}
\frac{1}{r}{\bar t}_4^{\infty}(q)=t_B^{\infty}(q)+t_{\Phi}^{\infty}(q)+ t_T^{\infty}(q)
\end{eqnarray}
where the three terms in the right-hand side are defined respectively by Eq.(\ref{t4Binf1}), Eq.(\ref{t4Phiinf}) and Eq.(\ref{t4t4inf}).

\begin{figure}
\centering
{\includegraphics[width=\linewidth]{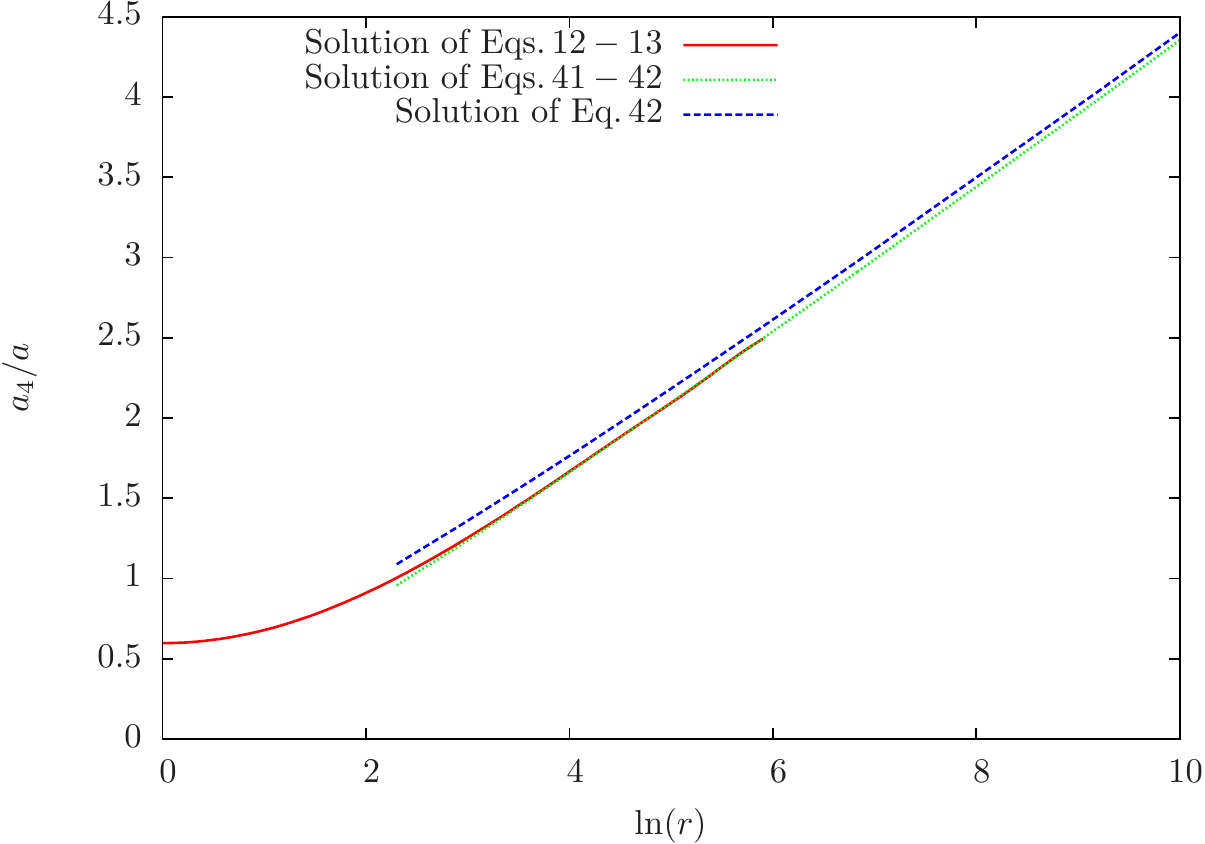}}
\caption{(Color online) Dimer-dimer scattering length $a_4$ as a function of the mass ratio $r$ (logarithmic scale).
The red full line is the exact numerical result obtained from Eq.(\ref{phitot}) and Eq.(\ref{t4tot}).
The green dotted line is the numerical result obtained from Eq.(\ref{phitotinf}) and Eq.(\ref{t4totinf}).
The blue dashed line is the result obtained from Eq.(\ref{t4totinf}) by setting $\Phi^{\infty}({\bf q}_1,{\bf q}_2)=0$.}
\label{fig3}
\end{figure}

Although these equations correspond to a very important simplification with respect to the original ones, they are still too
complicated to be solved analytically as such. In the following section we will show that they can be further simplified,
leading to an analytical answer. However it is of interest to solve them numerically to obtain the scattering length $a_4$
and to compare the result to the exact numerical solution of the original equations Eq.(\ref{phitot}) and Eq.(\ref{t4tot}).
This is done in Fig.~\ref{fig3}. It is quite surprising to see that, already for a mass ratio slightly above 10, the result from these asymptotic
equations coincide with the exact one within numerical precision. This provides naturally a further validation of these asymptotic
equations.

Finally let us consider the sum rule Eq.(\ref{sumrule}) in this limit. We have to handle the $|T_2|^2$ factor carefully, as we have
done just above. This implies in the same way that only $t_4(\textbf{Q},\Omega=0)=0$ appears in the sum rule and we end up with
the very simple relation:
\begin{eqnarray}\label{sumrule1}
\int_{0}^{\infty}\!\!dQ \;{\bar t}_4^{\infty}(Q)=1
\end{eqnarray}

\section{Discussion of the very heavy mass limit}\label{dvhma}

Although the equations have been much simplified in this heavy mass limit compared to the general ones,
they are still fairly complicated. We will nevertheless be able to come to a very simple conclusion by showing
that a quite natural hypothesis on the behaviour of ${\bar t}_4^{\infty}(Q)$ is fully consistent with the equations,
and is in agreement with results found numerically. However being able to prove that this is the only possible
solution looks a very difficult mathematical problem.

The natural hypothesis stems from the sum rule Eq.(\ref{sumrule1}) and from the fact that the scattering length
$a_4$ grows when the mass ratio increases, as it is known from preceding work \cite{pss,ism} and from our own
numerical calculations. It is natural to assume that $a_4$ grows indefinitely. From Eq.(\ref{a4barinf1}) this means
that ${\bar t}_4^{\infty}(0)$ grows indefinitely for large mass ratio. However the sum rule Eq.(\ref{sumrule1}) puts
a constraint. If we assume, as we have already done, that  ${\bar t}_4^{\infty}(Q)$ decreases rapidly when $Q$
is large and has a fairly regular behaviour, the increase of ${\bar t}_4^{\infty}(0)$, with fixed surface under the
curve ${\bar t}_4^{\infty}(Q)$ forced by the sum rule, implies that ${\bar t}_4^{\infty}(Q)$ becomes very peaked
around the origin for very large mass ratio.

In such a case we can further simplify the equations. Let us take first take $\Phi_T^{\infty}({\bf q}_1,{\bf q}_2)$
given by Eq.(\ref{phit4inf2}). Since ${\bar t}_4^{\infty}(Q)$ is peaked around the origin, $Q$ is actually forced
to be small. We can then expand the logarithm into $\ln [1+(Q+q_2)^2]/[1+(Q-q_2)^2] \simeq 
4Qq_2/(1+q_2^2)$. This leads to:
\begin{eqnarray}\label{phit4inf3}
\Phi_T^{\infty}({\bf q}_1,{\bf q}_2) \simeq \frac{8}{(1+q_1^2)(1+q_2^2)}   \int_{0}^{\infty} dQ\, {\bar t}_4^{\infty}(Q)=
\frac{8}{(1+q_1^2)(1+q_2^2)}
\end{eqnarray}
from the sum rule Eq.(\ref{sumrule1}). Hence $\Phi_T^{\infty}({\bf q}_1,{\bf q}_2)$ cancels exactly the Born 
contribution Eq.(\ref{phiborninf}). Therefore the only contributions left in the right-hand side of the $\Phi$ equation
are $\Phi_{\Phi 1}^{\infty}({\bf q}_1,{\bf q}_2)$ and $\Phi_{\Phi 2}^{\infty}({\bf q}_1,{\bf q}_2)$. However this
means that the $\Phi$ equation is now an homogeneous linear equation in $\Phi^{\infty}({\bf q}_1,{\bf q}_2)$,
without source term. Barring a singular kernel (which would make impossible in general to solve this $\Phi$
equation), the only solution is merely  $\Phi^{\infty}({\bf q}_1,{\bf q}_2)=0$.

Hence we are only left with the $t_4$ equation Eq.(\ref{t4totinf}) from which the $t_{\Phi}^{\infty}(q)$ term has
disappeared. In the equation Eq.(\ref{t4t4inf}) we can expand, for the same reason as above, the logarithm
into $\ln [4+(Q+q)^2]/[4+(Q-q)^2] \simeq 4Qq/(4+q^2)$ which gives:
\begin{eqnarray}\label{t4t4infinf}
t_T^{\infty}(q) \simeq -\frac{4}{\pi}\,\frac{1}{4+q^2}   \int_{0}^{\infty} dQ \,
{\bar t}_4^{\infty}(Q)=-\frac{4}{\pi}\,\frac{1}{4+q^2}
\end{eqnarray}
again from the sum rule. And we see again that this term $t_T^{\infty}(q)$ cancels exactly the Born term Eq.(\ref{t4Binf1}).
Hence the $t_4$ equation Eq.(\ref{t4totinf}) becomes merely ${\bar t}_4^{\infty}(q)/r=0$. But this is perfectly consistent
with the fact that we deal with $r=\infty$ limit. In conclusion we find that the equations are perfectly satisfied in
this limit by a function ${\bar t}_4^{\infty}(Q)$ very strongly peaked around the origin and satisfying the sum rule 
Eq.(\ref{sumrule1}). The only trouble is naturally that we have been unable to extract any information.
In order to obtain results we have to take more carefully advantage of the fact that ${\bar t}_4^{\infty}(Q)$ 
is strongly peaked.

For this purpose let us go back to the $\Phi$ equation Eq.(\ref{phitotinf}). The source term $\Phi_B^{\infty}({\bf q}_1,{\bf q}_2)
+\Phi_T^{\infty}({\bf q}_1,{\bf q}_2)$ can be factorized into:
\begin{eqnarray}\label{factsourc}
\Phi_B^{\infty}({\bf q}_1,{\bf q}_2)+\Phi_T^{\infty}({\bf q}_1,{\bf q}_2)=\frac{2}{(1+q_1^2)}\,S(q_2)
\end{eqnarray}
where:
\begin{eqnarray}\label{eqS}
S(q_2)=\frac{1}{q_2}   \int_{0}^{\infty} dQ\, \frac{{\bar t}_4^{\infty}
(Q)}{Q}\,\ln\frac{1+(Q+q_2)^2}{1+(Q-q_2)^2}-\frac{4}{1+q_2^2}
\end{eqnarray}
In the above discussion, valid for an extremely peaked function ${\bar t}_4^{\infty}(Q) \sim \delta(Q)$, we had $S(q_2)=0$.
When $r$ is large, but finite, ${\bar t}_4^{\infty}(Q)$ is peaked around the origin which makes $Q$ effectively bounded.
When $q_2$ is large, we can again expand the logarithm in  Eq.(\ref{eqS}) as above and reach again the conclusion
that $S(q_2)=0$. Hence $S(q_2)$ is also peaked around the origin.

Moreover, even if $S(q_2)$ is not zero, it retains the following exact property:
\begin{eqnarray}\label{Ssum}
 \int_{0}^{\infty}dk\,k^2\,S(k)=0
\end{eqnarray}
Indeed making use of the sum rule Eq.(\ref{sumrule1}) we can write from Eq.(\ref{eqS}):
\begin{eqnarray}\label{Ssum1}
 \int_{0}^{\infty}dk\,k^2\,S(k)=\int_{0}^{\infty}\!\!dQ \;{\bar t}_4^{\infty}(Q) \int_{0}^{\infty}\!\!dk \left[
 \frac{k}{Q}\,\ln\frac{1+(Q+k)^2}{1+(Q-k)^2}-\frac{4k^2}{1+k^2}\right]
\end{eqnarray}
The integral over $k$ can be calculated analytically, and it is found to be zero whatever the value of $Q$.

These two properties imply that the $\Phi$ equation has, to a very good approximation, a factorized solution
of the form:
\begin{eqnarray}\label{factPhi}
\Phi^{\infty}({\bf q}_1,{\bf q}_2)=F(q_1)S(q_2)
\end{eqnarray}
Indeed when this factorized expression is substituted in the $\Phi$ equation Eq.(\ref{phitotinf}) we see that
we can factorize $S(q_2)$ not only in the source term Eq.(\ref{factsourc}), but also in $\Phi_{\Phi 2}^{\infty}
({\bf q}_1,{\bf q}_2)$ (see Eq.(\ref{phiphi2inf})). In the remaining term $\Phi_{\Phi 1}^{\infty}({\bf q}_1,{\bf q}_2)$
given by Eq.(\ref{phiphi1inf}), when we substitute Eq.(\ref{factPhi}), the factor $S(k)$ prohibits large values for $|{\bf k}|$,
as we have just seen. But when $|{\bf k}|$ is small, or at most of order unity, it is a very good approximation to 
neglect it in the denominator $2 +q_1^2+({\bf q}_1+{\bf q}_2+{\bf k})^2$. Hence we are left in this term
with the integral $ \int d{\bf k}\,S(k)= 4\pi \int_{0}^{\infty}dk\,k^2\,S(k)=0$ as we have just seen in Eq.(\ref{Ssum}).
Therefore the contribution of the $\Phi_{\Phi 1}^{\infty}$ term is zero to a very good approximation, which shows
that the solution of the $\Phi$ equation is indeed of the form Eq.(\ref{factPhi}). We have checked that this decoupling
is properly satisfied asymptotically by the results of our numerical solution.

The resulting equation for $F(q)$ is:
\begin{eqnarray}\label{eqF}
F(q)=\frac{2}{1+q^2}-\frac{2}{\pi } \int_{0}^{\infty}dk\,\frac{k^2}{\sqrt{2+k^2}-1}\,\frac{F(k)}{2+k^2+q^2}
\end{eqnarray}
This equation is easily solved and gives a well behaved solution. This is most easily
understood if one notes that Eq.(\ref{eqF}) has an interesting physical interpretation. Indeed, as we
have seen, $\Phi({\bf q}_1,{\bf q}_2)$ describes the scattering of two free fermions on a dimer. In the
large mass ratio limit, it is reasonable to assume that the free heavy fermion does not play
any role and that only the scattering of the light fermion on the dimer is relevant. In this case
we are back to a problem first investigated by Skorniakov and Ter-Martirosian \cite{stm} for the
case of equal masses, and for which we have recently found an analytical solution in the case
of very different masses \cite{acl}. However there is a slight difference between our case and the
fermion-dimer scattering length problem. In this last one, since the kinetic energy is zero, the
total energy is just the dimer binding energy $-E_b$. In our case, since by definition of $\Phi$
the final state is made of two dimers with zero kinetic energy, the total energy is $-2E_b$,
that is twice the dimer binding energy. The two terms 2 present in the integral in the right-hand
side of Eq.(\ref{eqF}) can be tracked back to this total energy, taking into account that we have 
used reduced units. If we really had a fermion-dimer problem, these two 2 should be replaced by two 1.
In this case, making this substitution, we would rather have the equation:
\begin{eqnarray}\label{eqF1}
F'(q)=\frac{2}{1+q^2}-\frac{2}{\pi } \int_{0}^{\infty}dk\,\frac{k^2}{\sqrt{1+k^2}-1}\,\frac{F'(k)}{1+k^2+q^2}
\end{eqnarray}
Making the change $F'(q)=2 a_3(q)/(\sqrt{1+q^2}+1)$ we end up with:
\begin{eqnarray}\label{eqF1}
\frac{a_3(q)}{\sqrt{1+q^2}+1}=\frac{1}{1+q^2}-\frac{2}{\pi } \int_{0}^{\infty}dk\,\frac{a_3(k)}{1+k^2+q^2}
\end{eqnarray}
which is exactly the equation we had obtained \cite{acl} when one fermion in the dimer is very heavy while
the two other ones are very light. We have found the analytical solution $a_3(q)=1/(1+q^2)$.
This makes it easy to understand that the solution of Eq.(\ref{eqF}) is very similar and indeed we have found
the analytical solution:
\begin{eqnarray}\label{}
F(q)=\frac{4}{(1+q^2)(\sqrt{2+q^2}+1)}
\end{eqnarray}

This analysis makes it also possible to understand physically the factorization Eq.(\ref{factPhi}). Indeed since
the evolution of the heavy and of the light fermions decouples, it is quite natural that in the vertex
$\Phi$, which describes their evolution in the presence of a dimer, their contributions factorize
as it is the case for the wavefunction of two independent systems.

Having seen that the solution Eq.(\ref{factPhi}) for $\Phi$ is perfectly acceptable, we can substitute it
in the equation for $t_4$ Eq.(\ref{t4totinf}). However in the term $t_{\Phi}^{\infty}(q)$, given by Eq.(\ref{t4Phiinf}),
we can make use of the fact that we are only interested in small values of the variable $q$ since
we know that ${\bar t}_4^{\infty}(q)$ is peaked around the origin. In this case writing again
$\ln [1+(k'+q)^2]/[1+(k'-q)^2] \simeq 4k'q/(1+q^2)$ and substituting Eq.(\ref{factPhi}) leads to:
\begin{eqnarray}\label{t4Phiinfsimpl}
t_{\Phi}^{\infty}(q) = \frac{4}{\pi ^3}\,\frac{1}{1+q^2} \int_{0}^{\infty} dk\,
\frac{k^2}{1+k^2}\,\frac{F(k)}{\sqrt{2+k^2}-1} \int_{0}^{\infty} dk'\,k'^2\,S(k')=0
\end{eqnarray}
where the last equality comes again from the property Eq.(\ref{Ssum}).
Hence $\Phi$ disappears entirely from the equation for $t_4$ and we are left explicitly with:
\begin{eqnarray}\label{t4totinfsimpl}
\frac{1}{r}{\bar t}_4^{\infty}(q)=\frac{4}{\pi }\,\frac{1}{q^2+4}
-\frac{1}{\pi}\,\frac{1}{q}   \int_{0}^{\infty} dQ \,
\frac{{\bar t}_4^{\infty}(Q)}{Q}\,\ln\frac{4+(Q+q)^2}{4+(Q-q)^2}
\end{eqnarray}

We have checked numerically that taking $\Phi^{\infty}({\bf q}_1,{\bf q}_2)=0$ in Eq.(\ref{phitotinf})
and Eq.(\ref{t4totinf}) leads to the correct solution. The exact numerical solution of Eq.(\ref{t4totinf})
with $\Phi^{\infty}({\bf q}_1,{\bf q}_2)=0$ leads for $a_4$ to the results displayed in Fig.~\ref{fig3}.
We see that, for a mass ratio above 10, it gives a fairly good result for $a_4$, and that when increasing further
the mass ratio the result converges toward the exact one.
We note that the disappearance of $\Phi$ from the equation for $t_4$ means that the approximation
made by Pieri and Strinati \cite{ps} in their work on the dimer-dimer scattering length is fully valid
in the limit of large mass ratio. Indeed they made a ladder approximation, neglecting the possible
breaking of a dimer described by $\Phi$ and retaining only repeated scattering between the two dimers. 
We see that, since for the equal mass case, this approximation gives $a_4 \simeq 0.78$ instead of the exact 
\cite{pss1,bkkcl} $a_4 \simeq 0.60$ and since it becomes exact in the limit of very different masses, 
it turns out to be quite a good approximation for any value of the mass ratio.

If we perform as a final step the change of function ${\tilde t}_4^{\infty}(q)=\pi {\bar t}_4^{\infty}(q)$, Eq.(\ref{t4totinfsimpl}) becomes:
\begin{eqnarray}\label{t4fin}
\frac{1}{r}{\tilde t}_4^{\infty}(q)=\frac{1}{q^2+1}
-\frac{1}{2\pi}\,\frac{1}{q}   \int_{0}^{\infty} dQ \,
\frac{{\tilde t}_4^{\infty}(Q)}{Q}\,\ln\frac{1+(Q+q)^2}{1+(Q-q)^2}
\end{eqnarray}
This equation is identical to the one we had in the fermion-dimer problem \cite{acl} when we considered the large mass
ratio domain. The only apparent difference is that, in the left-hand side of the equation, the coefficient in \cite{acl} is $m\ds/
m\us$, while in the present case it is $1/r=m\us/m\ds$. However since in \cite{acl}, the considered limit is $m\ds/
m\us \to 0$ while we consider here $m\ds/ m\us \to \infty$, and since the scattering length $a_4$ is unchanged under
the exchange $m\us \leftrightarrow m\ds$, the equations are indeed identical. 

\begin{figure}
\centering
{\includegraphics[width=\linewidth]{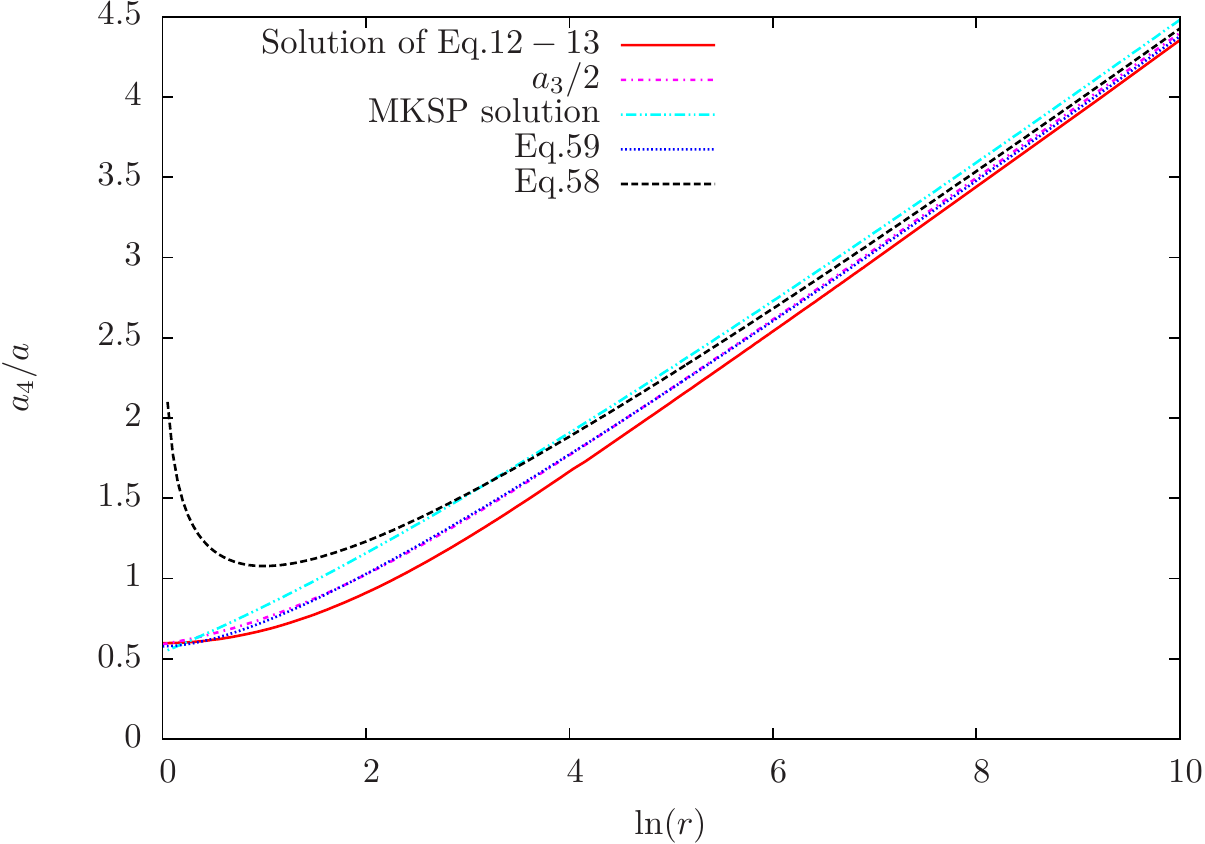}}
\caption{(Color online) Dimer-dimer scattering length $a_4$ as a function of the mass ratio $r$ (logarithmic scale).
The red full line is the exact numerical result obtained from Eq.(\ref{phitot}) and Eq.(\ref{t4tot}).
The purple dotted-dashed line is the exact numerical result for $a_3/2$.
The black dashed line is Eq.(\ref{a4fin})
The blue dotted line is Eq.(\ref{a4finmx}).
The light blue double-dotted-dashed line is the numerical solution of the equation $2 {\bar a}_4=2C+\ln(m\us/2 m\ds {\bar a}_4)$
found by MKSP \cite{mksp}.}
\label{asymp}
\end{figure}

Hence we have ${\tilde t}_4^{\infty}(0)={\bar a}_3=a_3/a$ in terms of the result obtained in \cite{acl}. Using
Eq.(\ref{a4barinf1}) and making the exchange $m\us \leftrightarrow m\ds$, we obtain our final result for the value 
of the dimer-dimer scattering length in this large mass ratio limit $m\us/ m\ds \to \infty$:
\begin{eqnarray}\label{a4fin}
a_4=\frac{a_3}{2}=\frac{a}{2}\bigg[ \ln (m\us /m\ds)-\ln (\ln (m\us /m\ds))+2C\bigg]
\end{eqnarray}
where $C=0.577215...$ is the Euler constant. On the other hand for $m\us = m\ds$ we have $a_4 \simeq 0.60$
while $a_3 \simeq 1.18$. Hence in this case we also have $a_4 \simeq a_3/2$. Hence it can be guessed that the equality
$a_4= a_3/2$ is approximatively valid whatever the mass ratio. Indeed the two quantities $a_4$ and $a_3/2$ 
are displayed in Fig.~\ref{asymp} and it can be seen that their difference is at most $\sim 0.1$, obtained for
$\ln(m\us /m\ds) \sim 2$. This difference is naturally expected since $a_4$ is invariant under $m\us \leftrightarrow m\ds$
while this is not the case for $a_3$.

The last expression in our result Eq.(\ref{a4fin}), which is displayed in Fig.~\ref{asymp}, is in agreement with the work of Marcelis, Kokkelmans, Shlyapnikov 
and Petrov (MKSP) \cite{mksp} for $a_4$ in this large $m\us /m\ds$ regime. They addressed this problem with the 4-body Schr\"odinger equation,
which they solved in this regime by a Born-Oppenheimer approximation. They found the approximate relation
$2 {\bar a}_4=2C+\ln(m\us/2 m\ds {\bar a}_4)$. To dominant order (i.e. omitting the $2 {\bar a}_4$ in the right-hand side)
it gives $2 {\bar a}_4=2C+\ln(m\us/m\ds )$ in agreement with two terms of our Eq.(\ref{a4fin}). Our third term is recovered
by inserting this expression in the right-hand side of their relation, corresponding to the next step in a recursive solution of this equation,
and keeping only the dominant contribution.

In this spirit one can find an analytical formula, slightly different from Eq.(\ref{a4fin}) but equivalent for large $m\us/ m\ds$,
which is much closer to the exact numerical result than our Eq.(\ref{a4fin}) or
also than the exact numerical solution (also displayed in Fig.~\ref{asymp}) of MKSP approximate relation. In Eq.(\ref{a4fin}) the trouble comes 
at low $m\us /m\ds$ from the divergence of $\ln (\ln (m\us /m\ds))$ when $m\us /m\ds \to 1$.
This is cured by introducing a constant in the logarithm. This constant could be adjusted for fine-tuning.
But taking it equal to 1 is both simple and in very good agreement with numerics. This leads to:
\begin{eqnarray}\label{a4finmx}
a_4=\frac{a}{2}\bigg[ \ln (m\us /m\ds)+2C-\ln [\ln (m\us /m\ds)+1]\bigg]
\end{eqnarray}
As it can seen from Fig.~\ref{asymp}, it is fairly close to the exact numerical result for $a_4$ (and even extremely
close to $a_3/2$). In general the agreement is expected to be limited by the fact that the exact result has the property to be
invariant under $m\us \leftrightarrow m\ds$ (which leads to a zero derivative with respect to $\ln (m\us /m\ds)$ for $m\us /m\ds=1$),
while an approximate analytical expression will not have this property. However we see that Eq.(\ref{a4finmx}) has precisely
a zero derivative with respect to $\ln (m\us /m\ds)$) for $m\us /m\ds=1$.

\section{Conclusion}

In this paper we have studied the dimer-dimer scattering length $a_4$ for a two-component Fermi mixture with different fermion
masses $m\us$ and $m\ds$ respectively. For this purpose we have made use of the exact field theoretical method already
present in the literature \cite{bkkcl,ism,lp}. The corresponding equations have been solved numerically for any value
of the mass ratio $m\us / m\ds$. However our main aim has been to study the large mass ratio domain. In this range
we have been able to simplify the equations enough to obtain an analytical solution. More specifically we have shown
that our final equation is essentially the same as the one obtained in the fermion-dimer scattering problem with
scattering length $a_3$. In this way we have shown that, for large mass ratio, we have the very simple result $a_4=a_3/2$.
Since this relation is also correct with a very good precision for $m\us=m\ds$, $a_4 \simeq a_3/2$ happens to be valid
for any mass ratio with a quite reasonable precision. We have found for the scattering length an analytical formula which
implements this finding and gives the result with a very good precision for any mass ratio. For the large mass ratio domain,
our result agrees for the dominant terms with the earlier one \cite{mksp} obtained from a study of the 4-body Schr\"odinger equation.

A very important finding in our investigation is that, in the large mass ratio domain, the dominant process in the dimer-dimer
scattering is the Born contribution, with resummation of any number of such processes. Other more complex processes, where 
for example one fermion propagates freely while the other one scatters on the other dimer, become negligible. Retaining
only these repeated dimer-dimer Born scatterings is precisely the approximation made by Pieri and Strinati \cite{ps} in their 
study of the dimer-dimer scattering. More precisely they made this approximate treatment in the case where the fermion
masses are equal $m\us=m\ds$. In this case there is no general justification for this approximation. Nevertheless it gives
for the scattering length a result $a_4 \simeq 0.78\,a$ which is not so far from the exact one $a_4 \simeq 0.60\,a$ (mostly if one
keeps in mind the simple Born result $a_4=2\,a$). As a consequence we come to the important conclusion that the Pieri and 
Strinati approximation is a fairly good one whatever the mass ratio. This is quite interesting since the processes they retain 
are much simpler than the ones which have to be considered in full generality.


\begin{references}
\bibitem{gps}For a review, see S. Giorgini, L. P. Pitaevskii and S. Stringari, Rev.Mod.Phys. {\bf 80}, 1215 (2008).
\bibitem{popov} V. N. Popov, Zh. Eksp. Teor. Phys. {\bf 50}, 1550 (1966), [Sov. Phys. JETP {\bf 23}, 1034 (1966)].
\bibitem{kk}L. V. Keldysh and A. N. Kozlov, Zh. Eksp. Teor. Phys. {\bf 54}, 978 (1968)
[Sov. Phys. JETP {\bf 27}, 521 (1968)]
\bibitem{eagles}D. M. Eagles, Phys. Rev. {\bf 186}, 456 (1969); D.M. Eagles, R.J. Tainsh, C. Andrikidis, Physica C {\bf 157}, 48 (1989).
\bibitem{leg}A. J. Leggett, J. Phys. (Paris), Colloq. {\bf 41}, C7-19 (1980); in \emph{Modern Trends in the Theory of
Condensed Matter}, edited by A. Pekalski and J. Przystawa (Springer, Berlin)
\bibitem{nsr}P. Nozi\`eres and S. Schmitt-Rink, J. LowTemp. Phys. {\bf 59}, 195 (1985).
\bibitem{haus}R. Haussmann, Z. Phys. B: Condens. Matter {\bf 91}, 291 (1993).
\bibitem{sdm}S.A.R. S\'a de Melo, M. Randeria and J.R. Engelbrecht, Phys.Rev.Lett. {\bf 71}, 3202 (1993).
\bibitem{ps}P. Pieri and G. C. Strinati, Phys. Rev. B {\bf 61}, 15370 (2000).
\bibitem{pss1}D. S. Petrov, C. Salomon, and G. V. Shlyapnikov, Phys. Rev. Lett. {\bf 93}, 090404 (2004).
\bibitem{pss}D. S. Petrov, C. Salomon, and G. V. Shlyapnikov, Phys. Rev. A {\bf 71}, 012708 (2005).
\bibitem{bkkcl}I.V. Brodsky, A.V. Klaptsov, M.Yu. Kagan, R. Combescot and X. Leyronas, J.E.T.P. Letters {\bf 82}, 273 (2005) and Phys. Rev. A {\bf 73}, 032724 (2006).
\bibitem{pss2}D. S. Petrov, C. Salomon, and G. V. Shlyapnikov, J. Phys. B: At. Mol. Opt. Phys. {\bf 38} S645 (2005);
Proceedings of the International School of Physics ÇÊEnrico FermiÊÈ Course CLXIV, Edited by M. Inguscio, W. Ketterle and C. Salomon : Ultracold Fermi Gases (Varenna, June 2006, IOS Press, Amsterdam 2008), p.385
\bibitem{ism}M. Iskin and C. A. R. S\'a de Melo, Phys. Rev. A {\bf 77}, 013625 (2008).
\bibitem{lp}J. Levinsen and D. S. Petrov, Eur. Phys. J. D {\bf 65}, 67 (2011).
\bibitem{acl}F. Alzetto, R. Combescot and  X. Leyronas, Phys. Rev. A {\bf 82}, 062706 (2010).
\bibitem{stm}G. V. Skorniakov and K. A. Ter-Martirosian, Zh. Eksp. Teor. Fiz. {\bf 31}, 775 (1956) [Sov. Phys. JETP {\bf 4}, 648 (1957)].
\bibitem{mksp}B. Marcelis, S. J. J. M. F. Kokkelmans, G. V. Shlyapnikov, and D. S. Petrov, Phys. Rev. A {\bf 77}, 032707 (2008).

\end{references}
\end{document}